# Imaging of the Coulomb driven quantum Hall edge states

Keji Lai[1,*], Worasom Kundhikanjana[1], Michael A. Kelly[1], Zhi-Xun Shen[1], Javad Shabani[2], and Mansour Shayegan[2]

[1]*Geballe Laboratory for Advanced Materials, Departments of Physics and Applied Physics, Stanford University, Stanford CA 94305*
[2]*Departments of Electrical Engineering, Princeton University, Princeton NJ 08544*

* E-mail: kejilai@stanford.edu

Abstract

**The edges of a two-dimensional electron gas (2DEG) in the quantum Hall effect (QHE) regime are divided into alternating metallic and insulating strips, with their widths determined by the energy gaps of the QHE states and the electrostatic Coulomb interaction. Local probing of these sub-micrometer features, however, is challenging due to the buried 2DEG structures. Using a newly developed microwave impedance microscope, we demonstrate the real-space conductivity mapping of the edge and bulk states. The sizes, positions, and field dependence of the edge strips around the sample perimeter agree quantitatively with the self-consistent electrostatic picture. The evolution of microwave images as a function of magnetic fields provides rich microscopic information around the $\nu = 2$ QHE state.**

The quantum Hall effect (QHE) is among the few textbook examples where the experimental results are insensitive to imperfections in real materials and solely determined by fundamental physics constants. After decades of research, the exact quantization of the Hall resistance in a two-dimensional electron gas (2DEG) system under strong magnetic ($B$) fields is now understood by the localization of electronic states when the bulk of the 2DEG is close to integer or fractional Landau level (LL) filling factors ($\nu$) [1]. Near the sample edges, however, the LLs bend up in energy due to the confining potential and intersect with the Fermi energy, resulting in alternating compressible (metal-like) and incompressible (insulator-like) strips [2-4]. In macroscopic samples, carriers propagating along the metallic edge channels are free from backscattering when scattered by impurities or inelastic events, therefore responsible for the topological robustness of the QHE [5]. The crucial role of edge states in the quantum Hall regime was recognized immediately after the proposal [2-5] and continues to attract research interest in recent years [6, 7].

Spatially resolved studies of the edge channels are usually challenging because most high mobility 2DEGs are located tens or even hundreds of nanometers below the surface of semiconductor heterostructures. Nevertheless, a number of novel designs, such as scanning gate microscopy [8-10], scanning single-electron transistor [11-13], and scanning charge accumulation microscopy [14-16], have shown compelling evidence of such edge modes by providing information on charge motion, surface potential, or local compressibility. Thorough studies of the local conductivity and the sizes of these edge channels, however, have not been achieved. In this letter, we demonstrate the conductivity mapping of the bulk and edge states in a GaAs/AlGaAs 2DEG using a cryogenic microwave impedance microscope (MIM) [17-19].

Narrow strips with either metallic or insulating screening properties are observed along edges of the sample as the system enters the QHE state. The evolution of the local conductivity distribution through the bulk filling factor $\nu_b = 2$ agrees with the self-consistent electrostatic calculation [3]. The imaging was performed without DC electrodes, vividly manifesting that the QHE edges are equilibrium states and do not depend on externally supplied currents.

The schematic setup of the variable-temperature ($T$) microwave microscope is shown in Fig. 1(a). An excitation power of 0.1 ~ 1μW at 1GHz is delivered to the shielded cantilever probe [20]. The reflected microwave is amplified by a cryogenic high electron mobility transistor (HEMT) amplifier and demodulated by a room-temperature quadrature mixer. The two output signals are directly proportional to the imaginary (MIM-Im) and real (MIM-Re) parts of the tip-sample admittance (inverse impedance) during the scan. The electronics in this experiment were set such that 1aF admittance change corresponds to 14mV in the output. The spatial resolution ~100nm is limited by the tip diameter rather than the wavelength of the microwave [17]. In order to create physical boundaries, the 2DEG sample was patterned into isolated dots, each with a diameter of 6~7μm. As shown in the atomic-force microscope line profile in Fig. 1(b), the 2DEG in the GaAs/AlGaAs interface (30nm below the surface) was etched away between the dots. The bulk electron density ($n_b = 3 \times 10^{11} cm^{-2}$) and mobility ($\mu = 5 \times 10^5 cm^2/Vs$) at $T = 2K$ were measured by DC transport on an unpatterned piece from the same wafer. We note that only the local diagonal conductivity $\sigma_{xx}$ is responsible for screening the in-plane radial microwave electric fields from the tip. The tangential current proportional to the Hall conductivity $\sigma_{xy}$ is irrelevant since it does not contribute to the screening.

The origin of quantum Hall edge states is strictly quantum-mechanical in nature. A semiclassical toy model, which intuitively suggests a conducting edge due to the cycloidal "skipping-orbit" motion, completely misses the essential physics of QHE. The non-interacting one-electron picture is also inadequate here because it leads to abrupt changes in density, prohibited by strong Coulomb penalty, where the Fermi level crosses a LL. When the electrostatic interaction is included [3], the density in real devices is depleted to zero near the sample edge by the confining potential, and rises smoothly toward $n_b$ with a length scale determined by the depletion width ($L$). The Landau quantization $\varepsilon_N = (N + 1/2)\, \hbar\omega_C$, where $N$ is the LL index and $\hbar\omega_C$ the cyclotron energy, gives rise to narrow constant-density regions with integer $\nu$'s. These highly resistive strips subdivide the edge into regions of different LL occupancy, commonly referred to as "edge states". The above scenario, including both the density profile and the energy diagram, is depicted in Figs. 2(a) and 2(b) using the actual sample parameters at $\nu_b = 2.31$. The depletion width $L = \varepsilon V_G/\pi n_b e \sim 110$nm sets the density profile at the edge [3], where $\varepsilon$ is the dielectric constant of GaAs, $V_G$ the band gap, and $e$ the electron charge. The $N = 0$ incompressible strip, which scales with $(a_B L)^{1/2}$ and $a_B \sim 10$nm being the effective Bohr radius in GaAs, is narrower than the compressible edge, whose width scales with $L$. Due to the small spin splitting in GaAs, each LL is 2-fold degenerate at this temperature so the $\nu = 1$ incompressible strip is ignored.

Figs. 2(c) and 2(d) show the MIM images at $\nu_b = 2.31$ ($B = 5.4$T) and $T = 2.3$K, with a typical line cut plotted in Fig. 2(e). In the extreme near-field regime, the tip-sample interaction is quasi-static and the impedance changes as a function of local $\sigma_{xx}$ can be computed by the finite-element analysis [18-20], as shown in Fig. 2(f). As detailed in the Supplementary Materials, the MIM response is a weighted average of the complex dielectric constant in a volume probed by the RF electric fields, which localize well underneath the tip for conducting 2DEG and extend up to several hundred nanometers for insulating 2DEG. We can therefore use simple 2D axisymmetric simulation to interpret the data for the wide etched region, the metallic edge, and the bulk. For an insulating strip as narrow as ~100nm sandwiched between conducting regions, the full 3D modeling is required. Using Fig. 2(f) as a guide, the non-monotonic conductivity distribution near the 2DEG edge is readily captured. First, the halo and the dark border near the physical boundary of the dot in Fig. 2(c) are topographic artifacts as the tip approaches and climbs up the 40nm step edge (Supplementary Materials). The effect is less problematic when the tip moves toward the interior for a distance close to the tip size, which coincides with the nominal depletion width. The MIM-Im signal then rises to a high value and stays for ~300nm before dropping slightly into the bulk [21]. The high MIM-Im and low MIM-Re signals here indicate a high local $\sigma_{xx} > 1\times10^{-4}$ $\Omega^{-1}$ of this band. The bulk conductivity ~$1\times10^{-5}$ $\Omega^{-1}$ is also determined by the lower MIM-Im and slightly higher MIM-Re signals than the metallic edge. Interestingly, in between these two regions, a narrow bright strip appears in the MIM-Re image, which can only be explained by the presence of a highly resistive channel with $\sigma_{xx}$ in the order of $10^{-7} \sim 10^{-8}$ $\Omega^{-1}$ [14]. This feature, which is also confirmed by 3D simulation with the tip scanning across a strip with fixed $\sigma_{xx}$, is not well resolved at higher $T$ or near $\nu_b = 4$ (Supplementary Materials), presumably due to the lower resistivity of the strip under those conditions. Using standard edge detection schemes, boundaries of different regions are determined by the midpoints of the rising and falling edges, e.g., arrows in Fig. 2(e). We then construct an idealized conductivity map in Fig. 2(g), which vividly demonstrates the non-trivial physics of the QHE edge states.

The microwave images [22] around $\nu_b = 2$ are shown in Figs. 3(b)–3(l), with the corresponding B-fields labeled on the transport data [Fig. 3(a)]. The conducting edge in MIM-Im and the resistive strip in MIM-Re are visible at $\nu_b = 2.60$ [Fig. 3(b)] and grow in width toward $\nu_b = 2$ [Figs. 3(c)–3(d)]. Discernible MIM-Re "patch" signals are seen at $\nu_b = 2.12$ [Fig. 3(e)] in the bulk, indicative of the decrease of bulk conductivity here. Near the integer QHE [Figs. 3(f)–3(i)], the conducting edge states are as wide as 1μm and the bulk electrons become highly inhomogeneous and inefficient to screen the microwave E-fields. We note that the bulk is not completely insulating, presumably due to the thermal excitation at 2.3K and the density fluctuation of the 2DEG sample. For increasing B [Figs. 3(j)–3(k)], electrons with local densities close to $n_{\nu=2}$ are still localized by disorder in the sample. Consequently, the center of the dot appears bright in the MIM-Re channel and dim in MIM-Im. As B further increases [Fig. 3(l)], the 2DEG delocalizes and regains the ability to screen E-fields, consistent with the homogeneous and metallic bulk revealed by the MIM. Such an evolution around $\nu_b = 2$ can be understood by schematics of the density profile across the dots depicted in Fig. 3(m), with a band of localized states continuously moving up in density as increasing B-fields.

In Fig. 4(a), we plot the widths of the metallic and insulating strips as a function of $\nu_b$, showing good agreements between the data and the electrostatic calculation [3] for $\nu_b > 2.2$. The deviation close to $\nu_b = 2$ may come from the density fluctuation that limits the edge channel widths. Such a quantitative comparison between theory and experiment has not been reported by other scanning techniques [8-16, 23, 24]. It is also clear that the width of the metallic edges cannot be estimated by the cyclotron diameter using the semiclassical skipping-orbit picture. Finally, we compare the transport $\sigma_{xx}(DC) = \rho_{xx} / (\rho_{xx}^2 + \rho_{xy}^2)$ and the bulk $\sigma_{xx}$ at 1GHz in Fig. 4(b). Toward $\nu_b = 2$, the bulk conductivity drops due to the reduction of carriers in the $N = 1$ LL. The decrease of $\sigma_{xx}$(1GHz), however, is more gradual than that of $\sigma_{xx}$(DC) measured by the voltage leads, which are in equilibrium with the edges. In fact, the plummet of transport resistance is mostly from the widening of the insulator-like strip, which drastically reduces the tunneling between the edge states and the bulk. For $|\nu_b - 2| < 0.1$, the local $\sigma_{xx}$(1GHz) extracted from the MIM images shows very large error bars, consistent with the significant inhomogeneity observed in the bulk. In macroscopic samples, small metallic puddles are decoupled from the edges and cannot induce backscattering. The transport is thus confined to the dissipationless edge channels, resulting in a vanishing $\rho_{xx}$ and a quantized $\rho_{xy}$ plateau [3, 4]. As the $B$ increases sufficiently above $\nu_b = 2$, the DC conductivity matches well with the bulk $\sigma_{xx}$ measured by MIM since the transport is now through the gradually delocalized 2DEG. We see that microscopically, entering and leaving an integer $N$ are different processes. And the exact integer filling factor does not necessarily occur in the middle of the quantized plateau.

In summary, we have directly imaged the quantum Hall edge channels using a microwave impedance microscope. The widths and local conductivity of both compressible and incompressible strips can be quantitatively compared with the electrostatic model. Our results pave the way to spatially resolve other exotic physics in 2DEG, such as the fractional QHE [1], the stripe and bubble phases [25], the 2D metal-insulator transition [26], and many more.

We would like to thank Steve A. Kivelson, Shoucheng Zhang, David Goldhaber-Gordon, and Bertrand I. Halperin for the helpful discussion. The work is supported by NSF grants DMR-0906027 and Center of Probing the Nanoscale PHY-0425897; DOE-DE-FG03-01ER45929-A001 for the equipment; NSF grants

**Figure captions:**

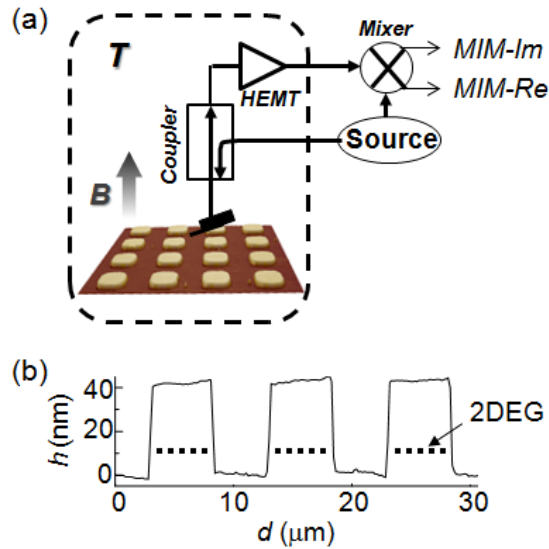

FIG. 1. (Color online) (a) Schematic setup of the microwave microscope and the 3D rendered image of the sample surface. The reflected 1GHz microwave from the cantilever tip is amplified and demodulated to form imaginary (MIM-Im) and real (MIM-Re) parts of the impedance maps. (b) A line profile of the surface topography through three dots. The 2DEG located 30nm below the surface is indicated in the plot.

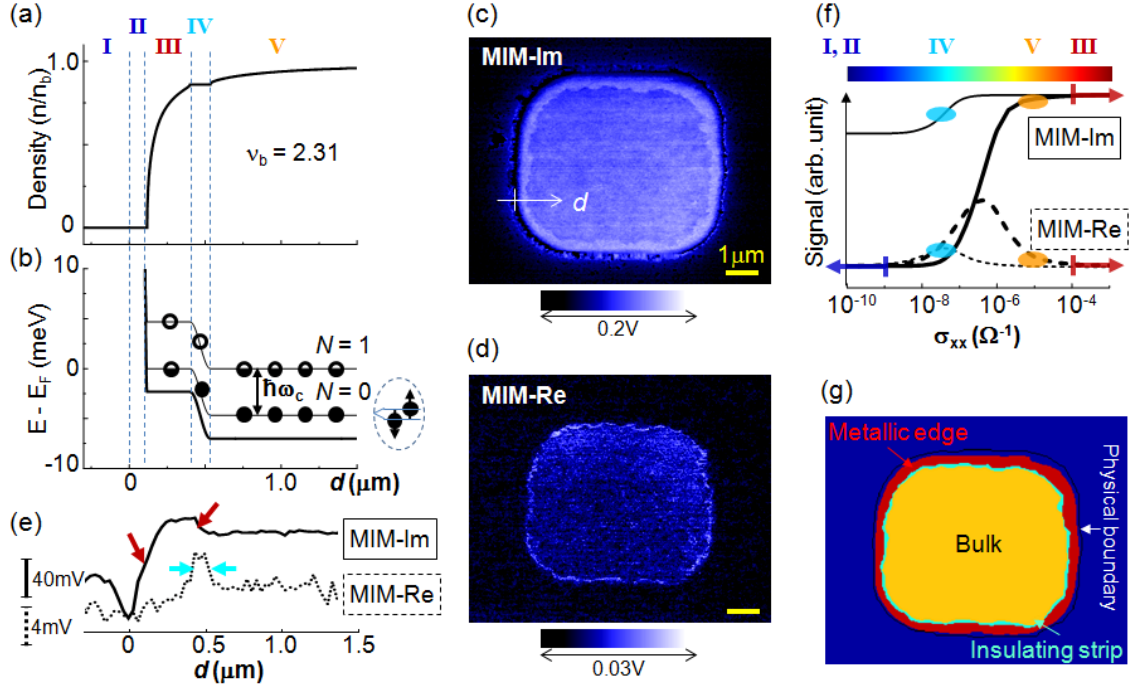

FIG. 2. (Color online) (a) Density profile and (b) energy diagram near the sample edge at the bulk filling factor $\nu_b$ = 2.31. The etched area (I), depletion region (II), metallic (III) and insulating (IV) strips, and the bulk (V) are labeled in the plot. The circles in the energy diagram (filled, half-filled, and empty) show the level occupancy. (c) MIM-Im and (d) MIM-Re images at $B$ = 5.4T and $T$ = 2.3K. The full color scale corresponds to 0.2V in MIM-Im and 0.03V in MIM-Re. The scale bars are 1μm. (e) Line cuts of the microwave data, labeled in (c). The vertical scales are 40mV for the MIM-Im (solid) and 4mV for the MIM-Re data (dashed). Rising and falling edges are indicated by arrows. (f) Results of the finite-element modeling, including the 2D axisymmetric analysis (thick solid and dashed lines) for the metallic edge and the bulk and the full 3D simulation (thin solid and dashed lines) for the insulating strip. (g) Idealized conductivity map combining the MIM images and the simulation.

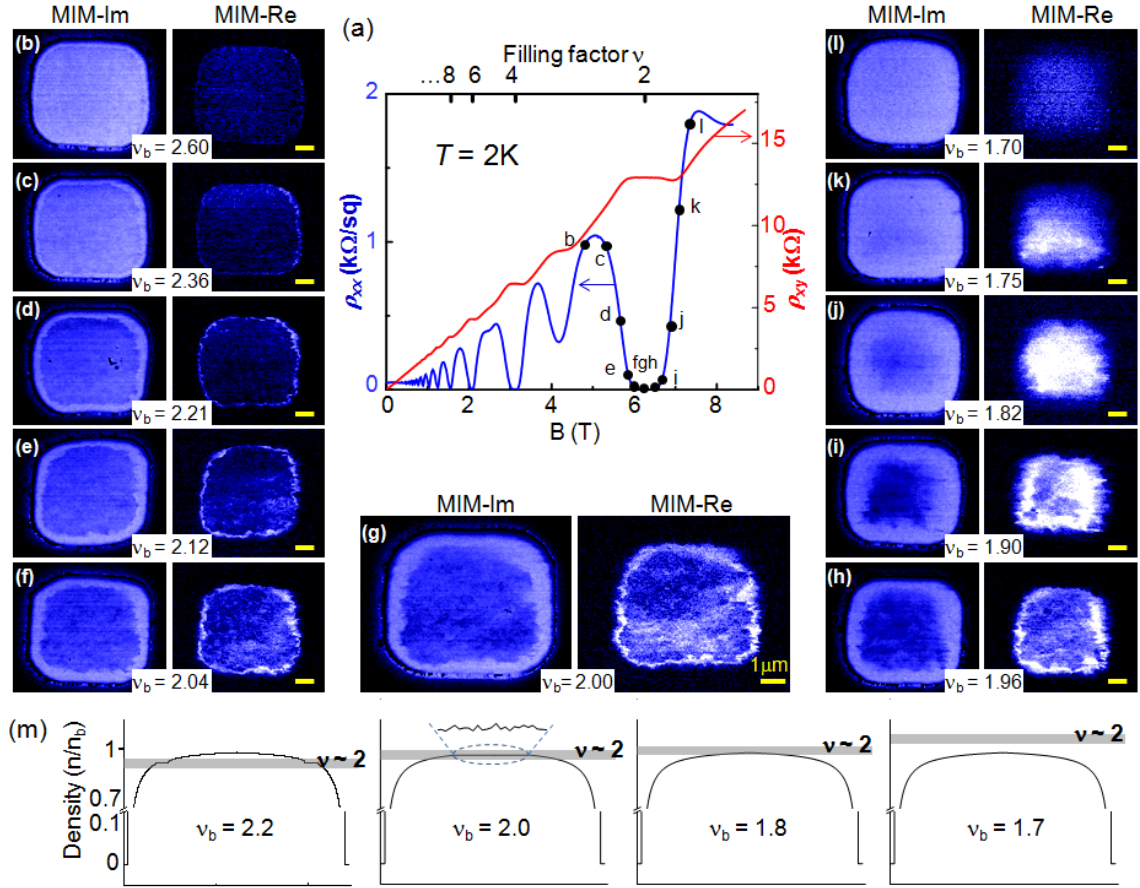

FIG. 3. (Color online) (a) Longitudinal and Hall resistivity as a function of $B$ or $\nu$ at 2K. The corresponding $B$-fields in (b – l) are labeled in the $\rho_{xx}$ trace. (b – l) Counterclockwise from top left to top right, MIM images at $T = 2.3$K as the $B$-field increases from 4.8T ($\nu_b = 2.6$) to 7.3T ($\nu_b = 1.7$). All scale bars are 1μm. The full color scales (not shown) are the same as Figs. 2(c) and 2(d). (m) From left to right, schematic density profiles across the center of the dots at $\nu_b = 2.2$, 2.0, 1.8, and 1.7, respectively. The shaded areas are sketches of the localized band.

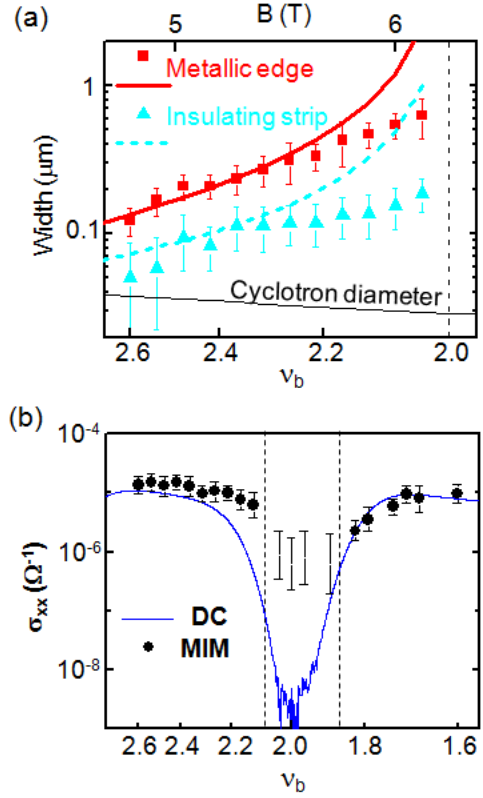

FIG. 4. (Color online) (a) Widths of the measured conducting edges (squares) and insulating strips (triangles). The solid and dashed lines are results from the electrostatic calculation [3]. The semiclassical cyclotron diameter with much smaller values and incorrect trend is also plotted for comparison. (b) Comparison between the macroscopic DC transport (solid line) and microscopic microwave (solid circles) conductivity. The $\sigma_{xx}$(1GHz) data between the two dashed lines show large uncertainties due to strong non-uniformity observed in the bulk of the 2DEG.

# Supplementary Materials

## Imaging of the Coulomb driven quantum Hall edge states

Keji Lai[1,*], Worasom Kundhikanjana[1], Michael A. Kelly[1], Zhi-Xun Shen[1], Javad Shabani[2], and Mansour Shayegan[2]

[1]*Geballe Laboratory for Advanced Materials, Departments of Physics and Applied Physics, Stanford University, Stanford CA 94305*
[2]*Departments of Electrical Engineering, Princeton University, Princeton NJ 08544*

*\*To whom correspondence should be addressed. E-mail: kejilai@stanford.edu.*

## A. 2D finite-element analysis (FEA) simulation

### Figure S1. Topographic artifact and the 2D axisymmetric modeling

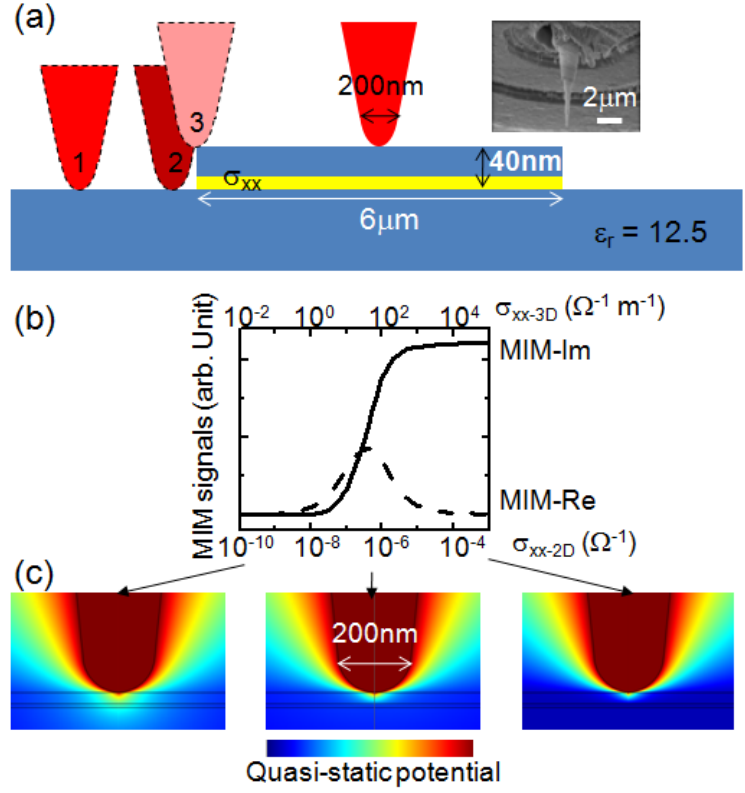

In Fig. 2(c), the halo outside the dot and the dark line right on the physical boundary are due to topographic artifacts. As shown in Fig. S1(a), the three "tips" with dashed boundaries on the left illustrate this effect due to capacitive coupling between the non-apex part of the tip and the 40nm step edge. The tip-sample capacitance increases from positions "1" to "2" because more material is seen to the right of the tip, and drops in position "3" because less material is seen to the left of the tip. The artifact diminishes as the tip moves toward the interior of the dot.

The tip-sample geometry for the simulation is sketched in Fig. S1(a). We assume the tip diameter $D$ = 200nm, consistent with the scanning electron micrograph (SEM) in the inset. The GaAs mesa is 6μm in diameter and 40nm above the etched area. The thickness of the 2DEG wave function is about $t$ = 10nm for a density of $3\times10^{11}$cm$^{-2}$ in the triangular quantum well. We have confirmed that since $t \ll D$, the results are unchanged as a function of the sheet conductance $\sigma_{xx-2D} = \sigma_{xx-3D} \cdot t$ for different $t$'s in the modeling.

We use a commercial FEA software COMSOL 3.5 in the AC/DC 2D axisymmetric mode. The software takes the geometry, generates a dense mesh, and solves for the quasi-static potential distribution. The imaginary (jωC) and real (1/R) components of the tip-sample admittance (inverse impedance), which are proportional to the MIM-Im and MIM-Re signals, are computed by the software through the two formula below.

$$\text{Energy} = \tfrac{1}{2} \int \varepsilon(\mathbf{r}) \cdot (\partial V / \partial \mathbf{r})^2 \, d\mathbf{r} = \tfrac{1}{2} CV^2 \qquad (1)$$

$$\text{Loss} = \tfrac{1}{2} \int \sigma(\mathbf{r}) \cdot (\partial V / \partial \mathbf{r})^2 \, d\mathbf{r} = \tfrac{1}{2} V^2 / R \qquad (2)$$

Fig. S1(b) shows the FEA results as a function of $\sigma_{xx}$ and Fig. S1(c) shows the corresponding potential distributions in the insulating limit ($\sigma_{xx} < 10^{-9} \Omega^{-1}$, left), the crossover ($\sigma_{xx} \sim 10^{-6} \Omega^{-1}$, middle), and the conducting limit ($\sigma_{xx} > 10^{-4} \Omega^{-1}$, right). The qualitative characteristics of the response can be understood as follows. When the 2DEG is highly resistive, the tip-sample interaction is mostly capacitive with no contrast compared with the etched regions. Toward the conducting limit, the 2DEG becomes the ground plane and the overall impedance is again lossless, with larger tip-ground capacitance than that of the etched regions. In between these two limits, the 2DEG resistance competes with the capacitance in the environment, resulting in increasing MIM-Im and non-monotonic MIM-Re signals as the sheet resistance decreases. Strictly speaking, the simulation only accounts for the microwave response in the center of the dot. However, as clearly seen in Fig. S1(c), the potential gradient (or the E-field) extends at most up to 3~5 tip diameters into the sample for low $\sigma_{xx}$ and well localizes underneath the tip toward the conducting limit, as is the case for both the metallic edge and the bulk. We therefore do not have to perform the much more complicated 3D simulation to understand the MIM signals in these two regions.

## B. 3D FEA simulation

**Figure S2. Geometry and results of the 3D modeling**

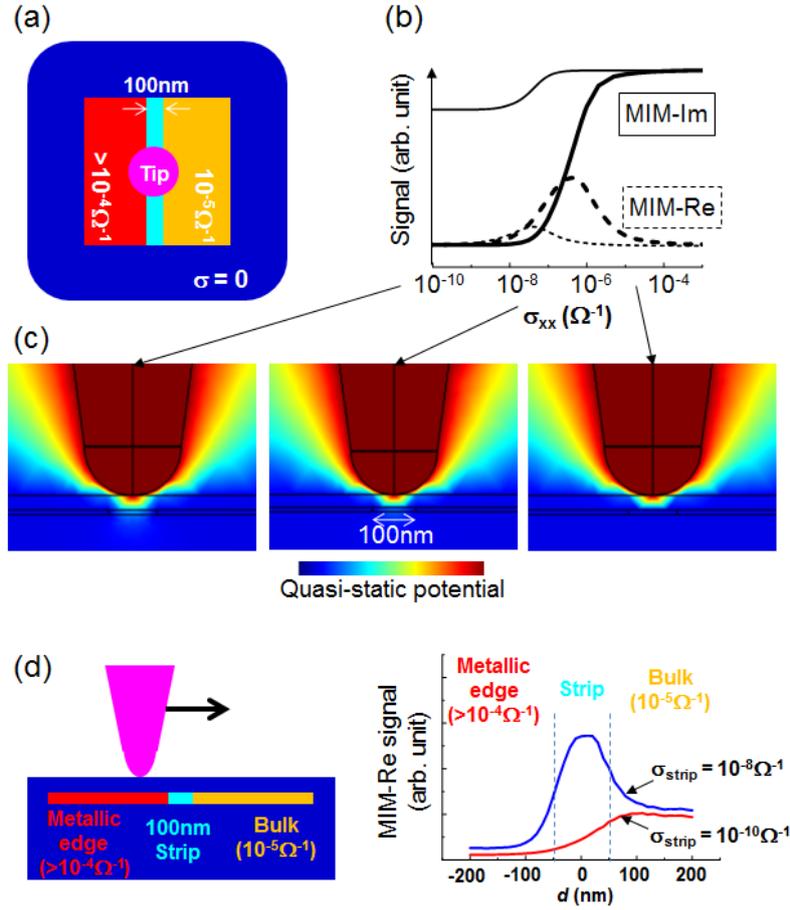

Because of the narrow width and the small $\sigma_{xx}$ of the incompressible strip, the 2D axisymmetric modeling in Fig. S1 is no longer valid here. The more complicated 3D FEA has to be utilized to simulate the tip-sample admittance with metallic regions (>$10^{-4}\Omega^{-1}$ for the metallic edge and $10^{-5}\Omega^{-1}$ for the bulk) in proximity, as schematically shown in Fig. S2(a). Here we take a typical strip width of 100nm and the results are plotted as the thin solid and dashed lines in Fig. S2(b). The cross-sectional views of the quasi-static potential distribution are shown here for three representative $\sigma_{xx}$'s of the strip for illustration of the underlying physics. The RF electric fields are squeezed inside the strip because they cannot enter the metallic domains on both sides. The effectively probed area is thus confined inside the strip with a length comparable to the tip diameter, resulting in an MIM-Re signal roughly proportional to the strip width. The loss peak here is also shifted to lower conductivity compared with the bulk case due to the stronger E-field in the strip [see equation (2)]. We also note that while there might be internal features across the insulating strip, the spatial resolution is relatively poor here. The fact that the strip conductivity matches within an order of magnitude to the peak position in Fig. 2(f) is also supported by other experiment, e.g., Ref. 14 in the main text. Therefore, it is reasonable to approximate the width of the insulating strip as the full-width-half-maximum of the MIM-Re peak.

A more intuitive way to understand the MIM-Re peak is to directly simulate the tip scanning across a 100-nm strip with fixed $\sigma_{xx}$, as schematically shown in Fig. S2(d). As is clearly seen, for an insulating strip with $\sigma_{xx} \sim 10^{-10}\Omega^{-1}$, there is no measurable peak in the MIM-Re channel. For the same strip with $\sigma_{xx} \sim 10^{-8}\Omega^{-1}$, however, a single peak with full-width-half-maximum of 100nm is obtained, again confirming the validity of our method to extract the strip width. The resemblance between the simulated signal and the data (line cut of MIM-Re in Fig. 2e) shows that one can indeed use Fig. 2f to identify the local conductivity of the strip.

**C. Dip in the MIM-Im channel corresponding to the insulating strip**

**Figure S3. Line cut taken at $\nu_b = 2.36$**

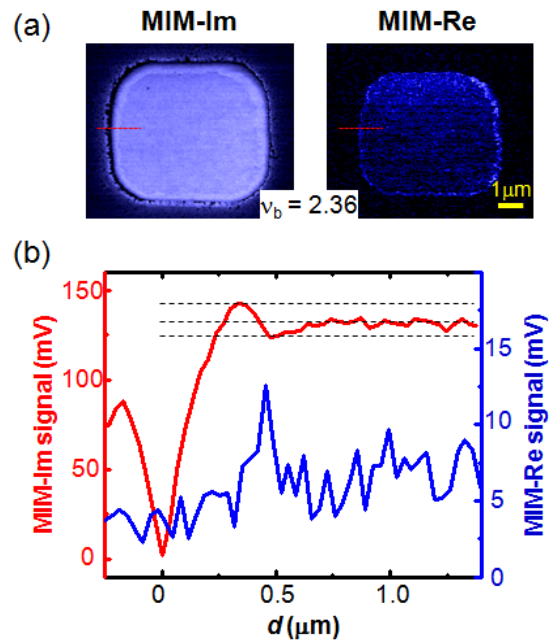

According to the 3D FEA in Fig. S2 and Fig. 2(f), there should be a small dip in the MIM-Im channel corresponding to the insulating strip. We indeed observe such features in some line cuts, with an example shown in Fig. S3 [same color scale as Figs. 2(c) and 2(d)]. This small dip, however, is easily smeared out by the signal fluctuation in the bulk due to either density inhomogeneity or measurement noise. We therefore do not identify the insulating strip in the MIM-Im image, which is hard anyway because the color scale here is dominated by the much larger contrast between the 2DEG and the etched region. It is easier to observe this strip in the MIM-Re channel, in which signal vanishes in both conducting and insulating limits.

## D. Absence of the MIM-Re strip at higher *T* and lower *B*

**Figure S4. MIM images under other conditions**

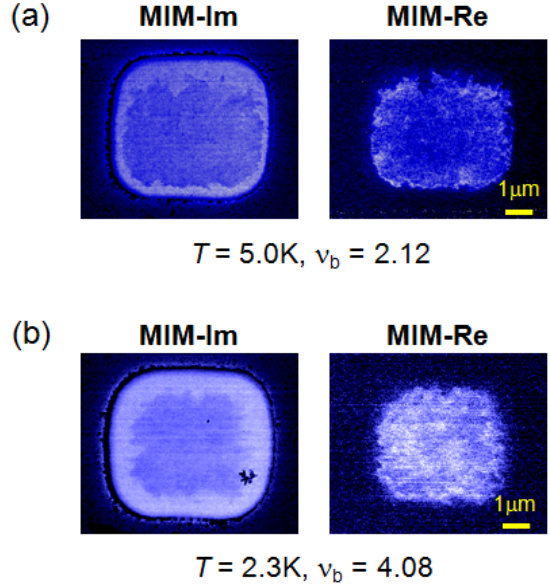

The observation of a narrow insulating strip in the MIM-Re implies that its local conductivity roughly aligns with the peak in Fig. 2(f), i.e., in the order of $10^{-7} \sim 10^{-8}$ $\Omega^{-1}$. The resistance of this strip may move out of this window as we vary the temperature or the magnetic field (filling factor), resulting in the absence of the MIM-Re peak (not necessarily the disappearance of the strip itself). Fig. S4(a) shows the MIM images taken near $\nu_b = 2$ but at higher $T = 5$K, and Fig. S4(b) at the same $T = 2.3$K but near $\nu_b = 4$. In both cases, while the metallic edge is still clearly seen in the MIM-Im channel, a resistive strip, if exists at all, is not well defined with respect to the bulk. Future work down to lower temperatures may help to elucidate this point through the subsequent appearance and disappearance in *T*-dependent experiments.

**E. Electrostatic picture and image analysis**

The basic results of the electrostatic picture in Ref. 3 [Phys. Rev. B 46, 4026 (1992)] is reproduced here (in SI rather than CGS units). Note that the half-width of the forbidden gap takes the place of the gate voltage in the original formula due to the pinning of the Fermi level by the surface states.

Depletion width $L$ $$L = 2l = \frac{\varepsilon V_G}{\pi n_b e} \tag{3}$$

Zero-field ($B = 0$) density profile $$n_{B=0}(x) = \left(\frac{x-l}{x+l}\right)^{1/2} n_b \tag{4}$$

Width of the $k^{\text{th}}$ compressible strip $$b_k = x_k - x_{k-1}, \quad x_k = l\frac{(v_b/2)^2 + k^2}{(v_b/2)^2 - k^2} \tag{5}$$

Width of the $k^{\text{th}}$ incompressible strip $$a_k = \sqrt{\frac{8\varepsilon\hbar\omega_c}{\pi e^2 (dn/dx)_{x=x_k}}} \tag{6}$$

Boundaries between different regions are determined by a standard image analysis algorism, Canny edge detection. In fact, different edge detection schemes yield similar results due to the well-defined rising and falling edges in the raw data, as indicated by arrows in Fig. 2(e). Once the boundaries are determined, for each pixel on the inner ring, the closest distance to the next ring is identified to calculate the edge channel widths in Fig. 4(a).